\documentclass[onecolumn,showpacs, preprint]{revtex4-1}
\usepackage{amsmath}
\usepackage{amssymb,ams fonts,times,natbib}
\usepackage{graphicx}
\usepackage{color}
\usepackage{subfig}
\usepackage{wrapfig}

\begin{document}

\title{Multidimensional  spectroscopy with a single broadband phase-shaped laser pulse}
\author{Rachel Glenn}
\author{Shaul Mukamel}
\affiliation{Department of Chemistry, University of California, Irvine, California 92697-2025, USA}
\date{\today}

\begin{abstract}
We calculate the
frequency-dispersed
nonlinear transmission signal of a phase-shaped visible
pulse to fourth order in the field. Two phase profiles, a phase-step and phase-pulse, are considered. Two dimensional signals obtained by varying the detected frequency and phase parameters are presented for a three electronic band model system. We demonstrate how two-photon and stimulated Raman  resonances can be manipulated by the phase profile and sign, and selected quantum pathways can be suppressed.

\end{abstract}

\maketitle

\section{Introduction}
Coherent control techniques, which utilize optimally shaped pulses to study the quantum interference and
 select quantum pathways
\cite{rice_new_1992,warren_coherent_1993,grumstrup_facile_2007,shim_how_2009}, have been widely used
to manipulate molecular stucture\cite{tannor_coherent_2007,shim_how_2009,lorenc_adaptive_2007,rice_optical_2000,bederson_advances_2001}, control chemical reactions\cite{assion_control_1998,lorenc_adaptive_2007,daniel_deciphering_2003} and
to infer the electronic and vibrational motions in molecules.
Pulse shaping techniques utilize the phase $\phi(\omega)$ of the field
\begin{equation}
\tilde{\mathcal{E}}(\omega)=\mathcal{E}(\omega)e^{i\phi(\omega)}
\label{singlepulse}
\end{equation}
to control the quantum pathways in matter. Typical choices for the phase profile are an
oscillating sinusoidal $\phi(\omega)=\alpha \sin(\omega-\omega )T$,
phase-step $\phi(\omega)=\theta(\omega-\omega_0)$, or chirp $\phi(\omega)=C(\omega-\omega_0)^2$
\cite{weiner_ultrafast_2011,goswami_optical_2003,shim_how_2009,Silberberg_book,lozovoy_systematic_2005}.
Another pulse shaping technique, which connects the time and frequency domain, is the use of a frequency comb\cite{eckstein_high-resolution_1978,femtosecond}. A frequency comb consists of a series of
evenly spaced pulses in the time domain and in the frequency domain the spectrum consists of sharp lines with
well defined frequencies.

Recent applications of phase control phase in one-dimensional spectroscopy  have been reported.
The spectral phase has been used to suppress processes such as two-photon absorption  (TPA),  by utilizing an asymmetric phase function\cite{meshulach_coherent_1999,Silberberg_2009,lozovoy_systematic_2005,
Dantus_2003,lozovoy_systematic_2005,zhang_control_2012} with respect to the TPA
transition frequency.  Shaped-pulses have been employed in Raman spectroscopy\cite{Silberberg_2009, Silberberg_book}.
A phase-step, in
coherent anti-Stokes Raman spectroscopy
(CARS), can significantly improve
the resolution  and reduce the non-resonant background\cite{Silberberg_2002_2, polack_control_2005, lim_single-pulse_2005, oron_all-optical_2004,roy_single-beam_2009,Silberberg_2009}.
By applying a non-abrupt phase-step, on the pump pulse in CARS,
 Raman resonances are narrower than compared to a transform limited pulse\cite{oron_narrow-band_2002}.
This technique\cite{postma_application_2008} can be used
to extract the line-width of the vibrational transitions in a molecule.
Pulse-shaping utilizing a oscillating phase in CARS, allowed the Raman spectrum to be extracted with high resolution and  relatively small background\cite{Silberberg_2002}.
By applying a narrow phase-pulse, a $\pi$-gate,
in single-beam CARS, the vibrational energy levels were mapped in a single measurement\cite{Silberberg_2002_3}.
Note that
Ref. \cite{Silberberg_2002_3} used an abrupt phase-pulse, where here we study a non-abrupt phase-pulse.

Coherent control is carried out using an  adaptive (closed loop)
pulse shaping scheme that employs genetic algorithm
 to optimize many control parameters\cite{nuernberger_femtosecond_2007,Silberberg_2010,
levis_selective_2001,baumert_femtosecond_1997,montgomery_general_2007,hornung_optimal_2000}  of    the laser pulses\cite{teaching_1992}.
Here we use a few control paramters.

\begin{figure}[t]
\begin{center}
\includegraphics[scale=0.3]{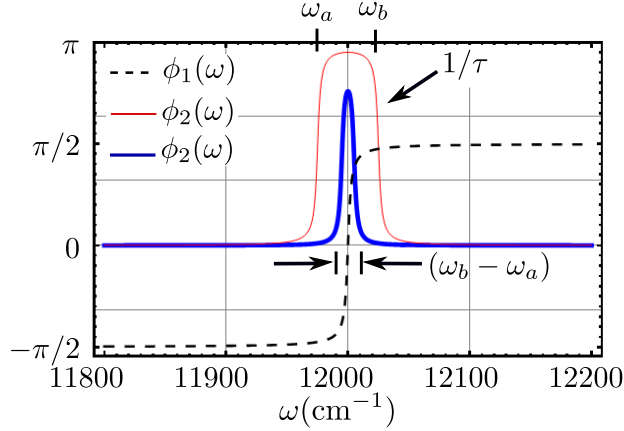}
\end{center}
\caption{(Color online)  The spectral phase $\phi(\omega)$ that we consider,
(dashed-black) is a phase step, Eq. \eqref{introarc} with transition width $1/\tau_a=2\mathrm{cm}^{-1}$;
(thick-blue) a narrow phase-pulse, Eq. \eqref{arctan}, with $\omega_a=11995\mathrm{cm}^{-1}$,
 $\omega_b=12005\mathrm{cm}^{-1}$, $\tau=0.5\mathrm{cm}$;
(thin-red) a wide phase-pulse, Eq. \eqref{arctan}, $\omega_a=11975\mathrm{cm}^{-1}$,
 $\omega_b=12025\mathrm{cm}^{-1}$, $\tau=0.5\mathrm{cm}$.
}
\label{fig:0}
\end{figure}

In this paper, we extend these applications to two-dimensional spectroscopy
by plotting the transmission spectrum of a broadband pulse
 as a function of the dispersed frequency
and the position of a  phase-step or a phase-pulse.
A phase-step and phase-pulse
 have been widely used in one-dimensional spectroscopy\cite{Silberberg_2002_2, polack_control_2005, lim_single-pulse_2005, oron_all-optical_2004,roy_single-beam_2009,postma_application_2008,oron_narrow-band_2002,Silberberg_2002,Silberberg_2009,bayer_2009,Silberberg_2010}.
 We investigate the control of the quantum pathways in the transmission of a single broadband pulse
 using the phase-pulse and phase-step shown in Fig. \ref{fig:0}.
The two-dimensional transmission signal for the phase-pulse and step
show diagonal peaks corresponding to two-photon absorption, Stokes processes, and Rayleigh processes. The two-photon absorption  and Stokes peaks are sensitive to the the sign
of the phase.

%
%
%
%
\section{The quartic transmission spectrum of a three-band model}
\begin{figure}[t]
\begin{center}
\includegraphics[scale=0.4]{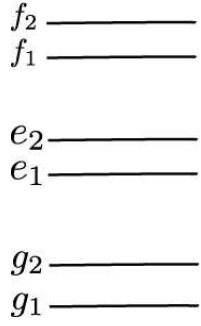}
\end{center}
\caption{(Color online)  The model level scheme, the corresponding
energy levels are
$g_1=0\mathrm{cm}^{-1}$,
$g_2=75\mathrm{cm}^{-1}$,
$e_1=12025\mathrm{cm}^{-1}$,
$e_2=12050\mathrm{cm}^{-1}$,
$f_1=24125\mathrm{cm}^{-1}$,
$f_2=24150\mathrm{cm}^{-1}$.
The dephasing rates were all chosen to be the same, $\gamma=10\mathrm{cm}^{-1}$.
}
\label{fig:level_scheme}
\end{figure}

%
%
%
%
We consider a three band system  (Fig. \ref{fig:level_scheme}) with electronic states $|g\rangle$, $|e\rangle$, $|f\rangle$
coupled to the
radiation field and described by the Hamiltonian
\begin{equation}
\hat{H}=\hat{H}_
e+\hat{H}_{int},
\end{equation}
where
\begin{equation}
\hat{H}_e=\sum_{\nu=g_i,\, e_i,\, f_i} \hbar \epsilon_\nu |\nu\rangle \langle \nu|,
\end{equation}
represents the system and $\hbar \epsilon _\nu$ is the energy of the state $\nu$.
The level scheme was chosen to highlight the resonances affected by the phase. $\hat{H}_{int}$ is the field-matter dipole interaction,
with the dipole operator $\hat{\mu}=\hat{V}^\dagger + \hat{V}$, where $V^\dagger$ $(V)$ is the matter raising and (lowering) operator, so that
$V(t)=\sum_e V_{ge}(t)|g\rangle \langle e| +\sum_f V_{ef}(t)|e\rangle \langle f| $
 and $V_{ij}(t)$ is the dipole matrix element in the interaction picture.
The classical electric field is $E(t)=\tilde{\mathcal{E}}(t)+\tilde{\mathcal{E}}^\dagger(t)$.
In the rotating wave approximation we have
\begin{equation}
\hat{H}_{int}=-\big (
\tilde{\mathcal{E}}^\dagger(t)  V(t)
+\tilde{\mathcal{E}}(t)  V^\dagger(t)
\big ).
\label{Hint}
\end{equation}
We shall calculate the frequency dispersed transmitted signal\cite{mukamelbook}
\begin{equation}
S(\omega)=-\frac{2}{\hbar}
\mathcal{I}
 \left[
\tilde{\mathcal{E}}^\dagger(\omega)
\int_{-\infty}^{\infty} dt \langle V_L (t) e^{-\frac{i}{\hbar}\int_{-\infty}^{\infty}H_{int-}(T)dT}\rangle e^{i\omega t}\right],
\label{tran}
\end{equation}
where $\mathcal{I} A(\omega)$ denotes the imaginary part of $A(\omega)$ and $\tilde{\mathcal{E}}^\dagger(\omega)$ is the Fourier transform:
$\tilde{\mathcal{E}}^\dagger(\omega)=\int_{-\infty}^{\infty}dt\mathcal{E}^\dagger(t)e^{i\omega t}$.
We use the superoperator formalism \cite{harbola}.
The superoperator $H_{int-}$ is defined as  $H_{int-}=H_{int\, L}-H_{int\, R}$.
The two superoperators $H_{int\, L}$ and $H_{int\, R}$ are defined by their actions
$H_{int\, L}X=H_{int}X$ and $H_{int\, R}X=XH_{int}$ \cite{harbola}.
The field is represented by its amplitude $\mathcal{E}(\omega)$ and phase $\phi(\omega)$,
Eq. \eqref{singlepulse}.

\begin{figure}[h!t]
\begin{center}
\includegraphics[scale=0.4]{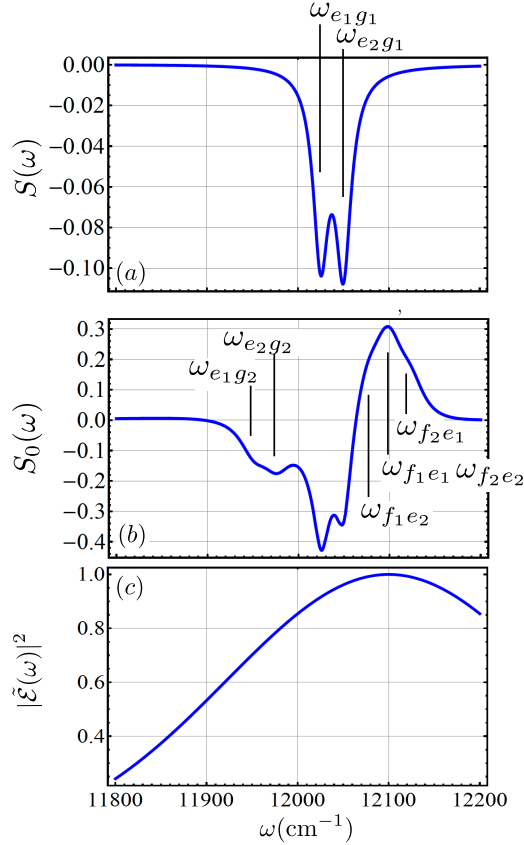}
\end{center}
\caption{(Color online)
(a) The linear absorption Eq.  \eqref{linear} of the model system of Fig. \ref{fig:level_scheme}.
(b)
The frequency dispersed transmission signal
$S_0(\omega)$, Eq. \eqref{S0}.
(c) The pulse power spectrum, $|\tilde{\mathcal{E}}(\omega)|^2$,  from Eq. \eqref{singlepulse}.
We used a Gaussian pulse Eq. \eqref{gaussian} with  $\phi=0$ for the field, Eq. \eqref{singlepulse},
$\sigma= 252 \mathrm{cm}^{-1}$ and $\Omega_1=12100 \mathrm{cm}^{-1}$.
}\label{fig:Linear}
\end{figure}

The linear absorption  spectrum is obtained by expanding Eq. \eqref{tran} to first-order
in $H_{int-}$
\begin{equation}
S(\omega)= -|\tilde{\mathcal{E}}(\omega)|^2  \mathcal{I} \,\chi^{(1)}(\omega),
\label{linear}
\end{equation}
where the linear susceptibility is
\begin{equation}
\chi^{(1)}(\omega)=\sum_{e_1 } -\frac{1}{\hbar}|\mu_{e_1 g_1}|^2G_{e_1 g_1}(\omega),
\end{equation}
   the Green's function is $G_{e_1g_1}(\omega)=(\omega-\omega_{e_1 g_1}+i\gamma)^{-1}$
and $\gamma$ is the dephasing rate.
 We use a Gaussian electric field
\begin{equation}
\mathcal{E}(\omega;\Omega_1,\sigma_1)=e^{-(\omega-\Omega_1)^2/(2\sigma_1^2)},
\label{gaussian}
\end{equation}
with center frequency $\Omega_1$ and standard deviation $\sigma_1$.
The pulse power spectrum, $|\tilde{\mathcal{E}}(\omega)|^2$, Eq. \eqref{singlepulse} with
 a Gaussian pulse, Eq. \eqref{gaussian}, is shown in  Fig. \ref{fig:Linear}(c).
Equation \eqref{linear} is plotted in Fig. \ref{fig:Linear}(a), with the dipole moments and $\hbar$ are set to one.
The two peaks in Fig. \ref{fig:Linear}(a) correspond to the transition frequencies $\omega_{e_1 g_1}$ and $\omega_{e_2 g_1}$.

\begin{figure}[t]
\begin{center}
\includegraphics[scale=0.5]{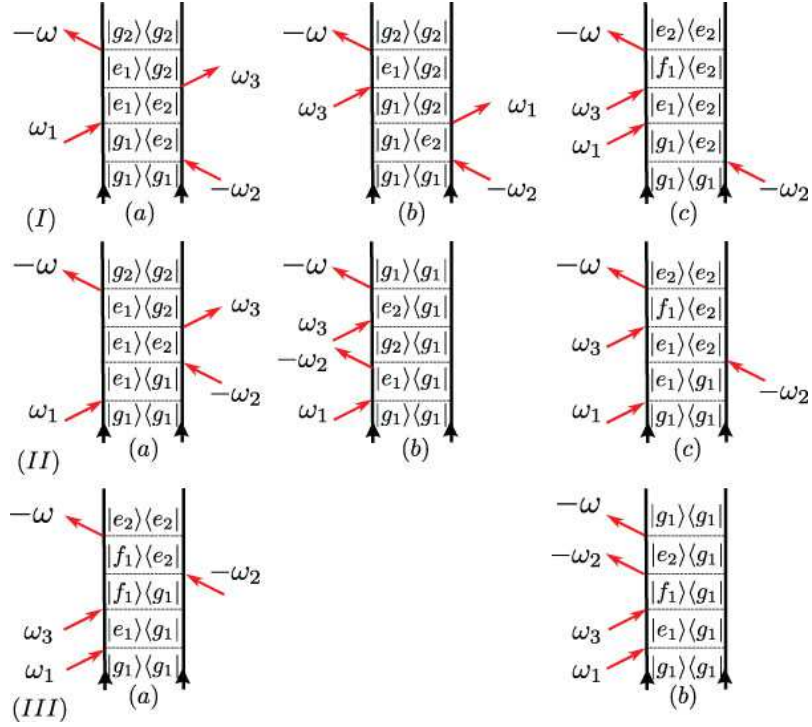}
\end{center}
\caption{(Color online) Ladder diagrams for the frequency dispersed transmitted signal
signal  Eq. \eqref{tran} expanded to third-order in $H_{int}$  from a single pulse.
The frequencies $\omega_1$ and $\omega_2$ correspond the variables
$\omega_1$, $\omega_2$ in the susceptibility, Eq. \eqref{overallS}.
 }
\label{fig:overall}
\end{figure}

The ladder diagram expansion of Eq. \eqref{tran} to third-order in $H_{int}$ is shown in Fig.  \ref{fig:overall}. The transmission signal\cite{mukamelbook} is given as
\begin{widetext}
\begin{eqnarray}
S(\omega)=&&
\frac{2}{\hbar}
\mathcal{I}
\int_{-\infty}^{\infty}\!\!
\int_{-\infty}^{\infty}\!\!\!
{d\omega_1}
{d\omega_2}
\tilde{\mathcal{E}}^*(\omega)
\tilde{\mathcal{E}}(\omega_1)
\tilde{\mathcal{E}}^*(\omega_2)
\tilde{\mathcal{E}}(\omega-\omega_1+\omega_2)
2 \pi  \chi^{(3)}(-\omega;-\omega_2,\omega_1, \omega-\omega_1+\omega_2).
\nonumber\\
\label{Si}
\end{eqnarray}
 The third-order susceptibility is
\begin{eqnarray}
&&
 \chi^{(3)}(-\omega;-\omega_2,\omega_1, \omega-\omega_1+\omega_2)
=
\chi^{(3)}_{I} (-\omega;-\omega_2,\omega_1, \omega-\omega_1+\omega_2)
\nonumber\\&&+
\chi^{(3)}_{II} (-\omega;-\omega_2,\omega_1, \omega-\omega_1+\omega_2)
+
\chi^{(3)}_{III} (-\omega;-\omega_2,\omega_1, \omega-\omega_1+\omega_2),
\label{overallS}
\end{eqnarray}
where the three terms correspond to diagrams (I), (II), (III) in Fig. \ref{fig:overall}.
\begin{eqnarray}
\chi_{I}^{(3)}
(-\omega;,-\omega_2,\omega_1,\omega-\omega_1+\omega_2)&&=
\left(\frac{-1}{2\pi \hbar}\right)^3
\hspace{-0.15in}
\sum_{e_1, e_2, g_2, f_1}
\hspace{-0.15in}
V_{g_1 e_2}
V_{ g_1 e_1}^*
V_{  e_2 g_2}^*
V_{ e_1 g_2}
G_{e_2 g_1}(\omega_2)
G_{e_1 e_2}(\omega_1-\omega_2)
G_{e_1 g_2}(\omega)
\nonumber\\&&
+
V_{g_1 e_2}
V_{  e_2 g_2}^*
V_{  g_1 e_1}^*
V_{ e_1 g_2}
G_{e_2 g_1}(\omega_2)
G_{g_1 g_2}(\omega_1-\omega_2)
G_{e_1 g_2}(\omega)
\nonumber\\&&
-
V_{g_1 e_2}
V_{ g_1 e_1}^*
V_{  e_1 f_1}^*
V_{ f_1 e_2}
G_{e_2 g_1}(\omega_2)
G_{e_1 e_2}(\omega_1-\omega_2)
G_{f_1 e_2}(\omega),
\label{S1}
\end{eqnarray}
\begin{eqnarray}
\chi_{II}^{(3)}
(-\omega;\omega_1,-\omega_2,\omega-\omega_1+\omega_2)&&=
\left(\frac{-1}{2\pi \hbar}\right)^3
\hspace{-0.15in}
\sum_{e_1, e_2, g_2, f_1}
\hspace{-0.15in}
V_{g_1 e_1}^*
V_{ g_1 e_2}
V_{  e_2 g_2}^*
V_{ e_1 g_2}
G_{e_1 g_1}(\omega_1)
G_{e_1 e_2}(\omega_1-\omega_2)
G_{e_1 g_2}(\omega)
\nonumber\\&&
+
V_{g_1 e_1}^*
V_{ e_1 g_2}
V_{  g_1 e_2}^*
V_{ e_2 g_1}
G_{e_1 g_1}(\omega_1)
G_{g_1 g_2}(\omega_1-\omega_2)
G_{e_2 g_1}(\omega)
\nonumber\\&&
-
V_{g_1 e_1}^*
V_{ g_1 e_2}
V_{  e_1 f_1}^*
V_{ f_1 e_2}
G_{e_1 g_1}(\omega_1)
G_{e_1 e_2}(\omega_1-\omega_2)
G_{f_1 e_2}(\omega),
\label{S2}
\end{eqnarray}
\begin{eqnarray}
\chi_{III}^{(3)}
(-\omega;\omega_1,\omega-\omega_1+\omega_2,-\omega_2)&&=
\left(\frac{-1}{2\pi \hbar}\right)^3
\hspace{-0.15in}
\sum_{e_1, e_2, g_2, f_1}
\hspace{-0.15in}
V_{g_1 e_1}^*
V_{ e_1 f_1}^*
V_{  f_1 e_2}
V_{ e_2 g_1}
G_{e_1 g_1}(\omega_1)
G_{f_1 g_1}(\omega+\omega_2)
G_{e_2 g_1}(\omega)
\nonumber\\&&
-
V_{g_1 e_1}^*
V_{ e_1 f_1}^*
V_{ g_1 e_2}
V_{ f_1 e_2}
G_{e_1 g_1}(\omega_1)
G_{f_1 g_1}(\omega+\omega_2)
G_{f_1 e_2}(\omega).
\label{S3}
\end{eqnarray}
\end{widetext}
An alternative form for the susceptibility  derived  using the loop diagrams is given in Appendix \ref{sec:loop}.   They represent the wavefunction in Hilbert-space,  instead of the density matrix, so that $\gamma$ in Hilbert
space represents the inverse lifetime.  Equations \eqref{App1}-\eqref{App4}, contain fewer terms to integrate compared to the Liouville-space. However, it is easier to perform the numerical integration in Eq. \eqref{Si},  using Mathematica in Liouville-space than in Hilbert-space.
For this reason, we use the Liouville-space  representation.

The total transmission spectrum of an unshaped transform limited pulse with $\phi=0$,
\begin{equation}
S_0(\omega)=S(\omega, \phi(\omega)=0),
\label{S0}
\end{equation}
is shown in Fig. \ref{fig:Linear}(b).
The transition frequencies are marked, based on the
 energy level diagram Fig. \ref{fig:level_scheme}.
The two  absorption peaks at
 $\omega=\omega_{e_1 g_1}$, $\omega_{e_2 g_1}$ in Figs.  \ref{fig:Linear}(b),
correspond to the Rayleigh process, where as,
$\omega_{e_1 g_2}$, $\omega_{e_2 g_2}$
represent the Stokes process.
 $\omega_{e_1 g_1}$, $\omega_{e_2 g_1}$  are the most pronounced;
they are also seen in the linear signal in Fig. \ref{fig:Linear}(a).
The four emission peaks at
$\omega=\omega_{f_1 e_1}$, $\omega_{f_2 e_1}$, $\omega_{f_1 e_2}$ and $\omega_{f_2 e_2}$
in Figs.  \ref{fig:Linear}(b)
correspond to two-photon absorption.
The $\omega_{f_1 e_2}$ peak
is difficult to distinguish due to the transition
from an absorption to an emission near that wavelength.
In Appendix B, we separate $S_0(\omega)$ into three
components
$S_0^I(\omega)$, $S_0^{II}(\omega)$, $S_0^{III}(\omega)$,
corresponding to $\chi^{(3)}_I$, $\chi^{(3)}_{II}$, $\chi^{(3)}_{III}$. These are
plotted in Fig. \ref{fig:appendix0}.

\begin{figure}[h!t]
\includegraphics[scale=0.65]{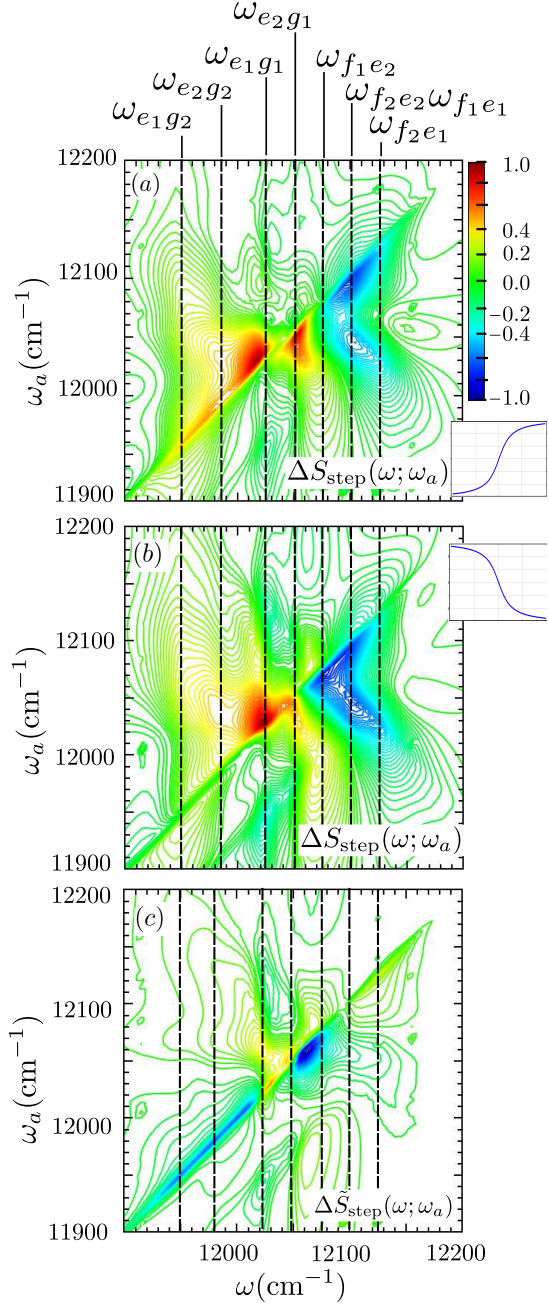}
\caption{(Color online) The two-dimensional  frequency dispersed transmission signal
$\Delta S_{\mathrm{step}}(\omega,\omega_a)$, Eq. \eqref{diff},
 with phase $\phi_1(\omega,\omega_a)$ is plotted; (a) $\tau_a=0.5 \mathrm{cm}$; (b) $\tau_a=-0.5 \mathrm{cm}$.
(c) The difference between (a) and (b), $\Delta \tilde{S}_{\mathrm{step}}(\omega,\omega_a)$,
 Eq. \eqref{posneg}.
The vertical dashed-black lines mark the transition frequencies.
The insets show the shape of the phase-step.
}
\label{fig:2Dintense}
\end{figure}

\begin{figure*}[t]
\begin{center}
\includegraphics[scale=0.3]{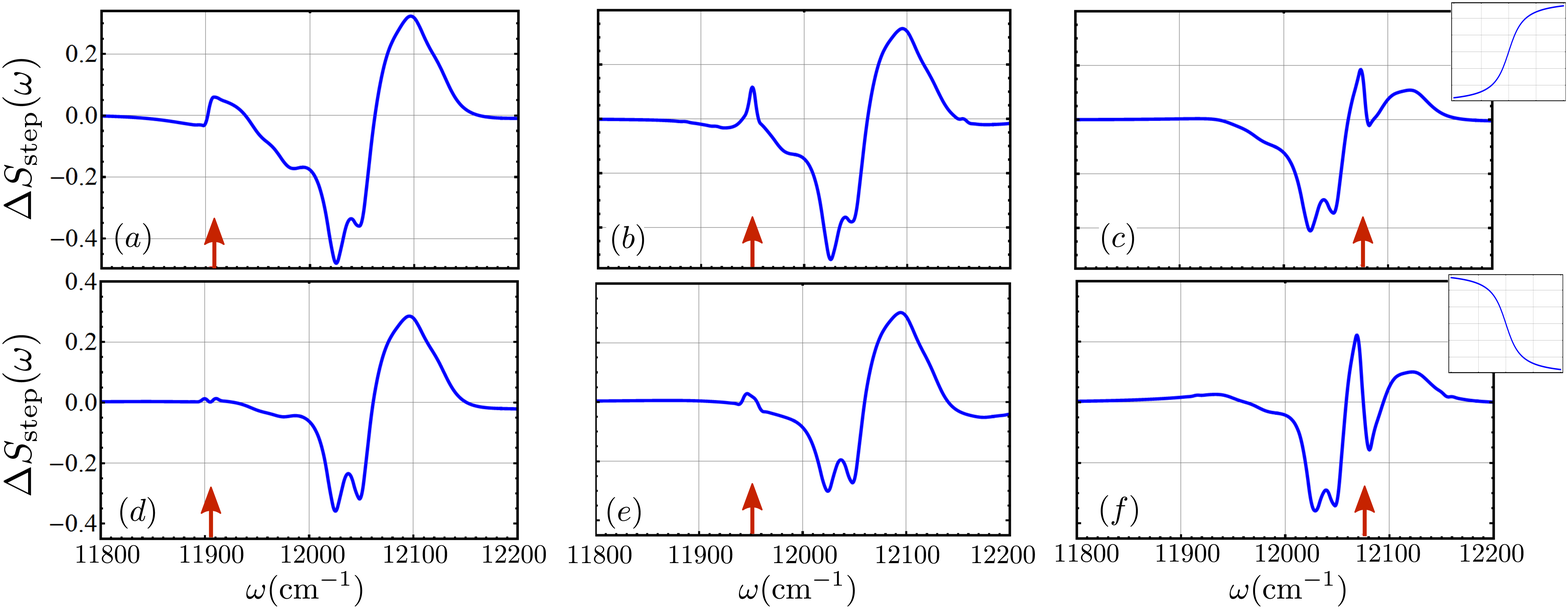}
\end{center}
\caption{(Color online) The transmission signal,  Eq. \eqref{Si},
using a phase-step, $\phi_1(\omega,\omega_a)$.
Top-row: positive-step, $\tau_a=0.5 \mathrm{cm}$, (a) $\omega_a=11905 \mathrm{cm}^{-1}$,
(b) $\omega_a=\omega_{e_2g_2}$, (c) $\omega_a=\omega_{f_1e_1}$.
Bottom-row: negative phase-step, $\tau_a=-0.5 \mathrm{cm}$, (d) $\omega_a=11905 \mathrm{cm}^{-1}$,
(e) $\omega_a=\omega_{e_2g_2}$, (f) $\omega_a=\omega_{f_1e_1}$. The red arrows
marks the position of the phase-step.
The insets show the shape of the phase-step.
}\label{fig:Diff_steps}
\end{figure*}

%
%
%
%
%
\section{ Two Dimensional nonlinear Transmission signal with a phase-step}
%
%
%

We first consider a $\pi$-phase-step phase
\begin{equation}
\phi_1(\omega)=\arctan[ \tau_a (\omega-\omega_a)],
\label{introarc}
\end{equation}
which has finite transition $\tau_a$, and position $\omega_a$, as marked in Fig. \ref{fig:0}.
The phase-step in Fig. \ref{fig:0} has transition width $1/\tau_a=2\mathrm{cm}^{-1}$.

The integrals in  Eq. \eqref{Si} were calculated numerically, with Gaussian pulses, Eq. \eqref{gaussian},  $\sigma_1=252\mathrm{cm}^{-1}$,  $\Omega_1=12100 \mathrm{cm}^{-1}$ and the dipole moments  set to one.

The two-dimensional  frequency dispersed transmission signal
\begin{equation}
\Delta S_{\mathrm{step}}(\omega;\omega_a)=
S(\omega; \phi_1(\omega,\omega_a,\tau_a))
-
S_0(\omega),
\label{diff}
\end{equation}
 is plotted in Fig. \ref{fig:2Dintense}(a) for a positive phase-step  $\tau_a=0.5 \mathrm{cm}$.
The vertical black-dashed  lines mark the positions of the
transition peaks.
\begin{figure}[h!t]
\includegraphics[scale=0.65]{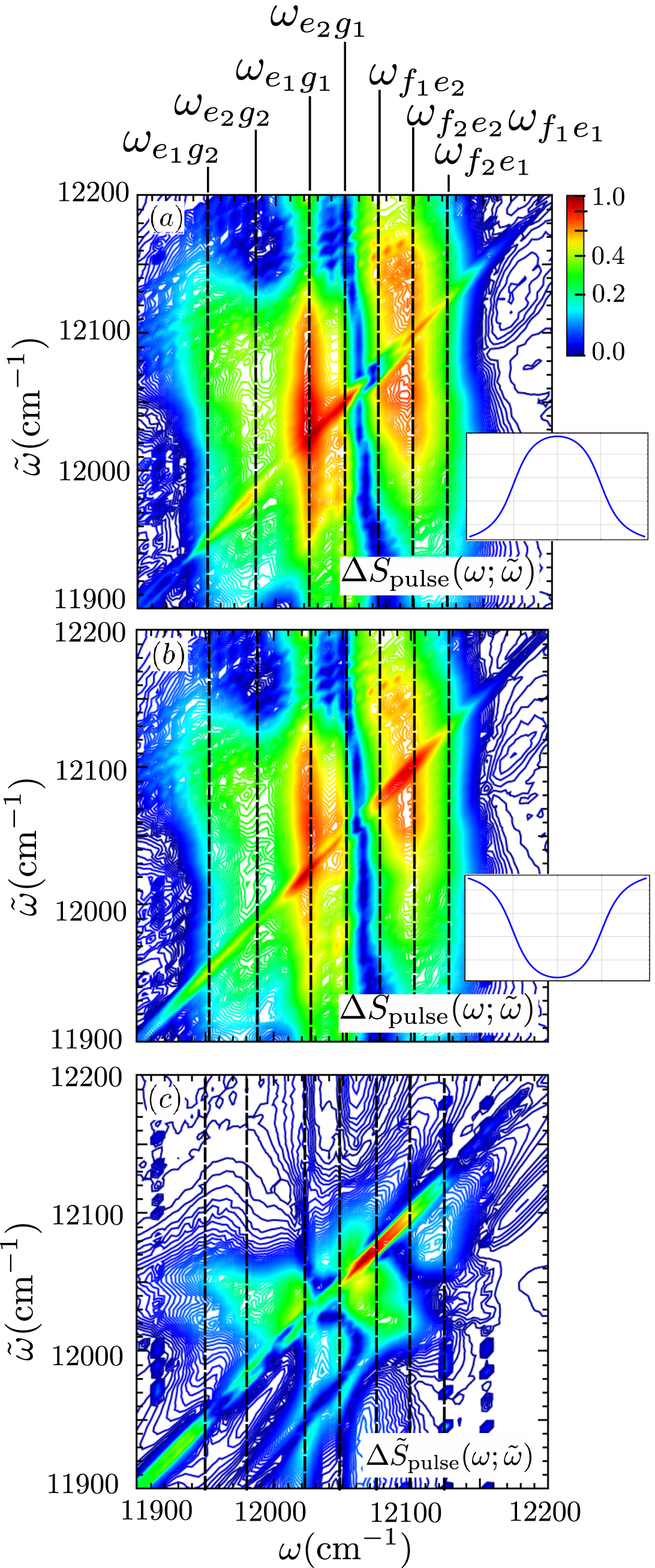}
\caption{(Color online)  The two-dimensional  frequency dispersed transmission signal Eq. \eqref{diffpulse}
is plotted with a narrow phase-pulse, $\phi_2(\omega,\tilde{\omega})$  width $\Delta\omega=10 \mathrm{cm}^{-1}$;
(a)  $\tau_a=0.5 \mathrm{cm}$, (b) $\tau_a=-0.5 \mathrm{cm}$.
(c) The difference between between (a), (b), Eq. \eqref{Diff2}.
The insets show the shape of the phase-pulse.
}\label{fig:Pulse}
\end{figure}
\begin{figure}[h!]
\begin{center}
\includegraphics[scale=0.6]{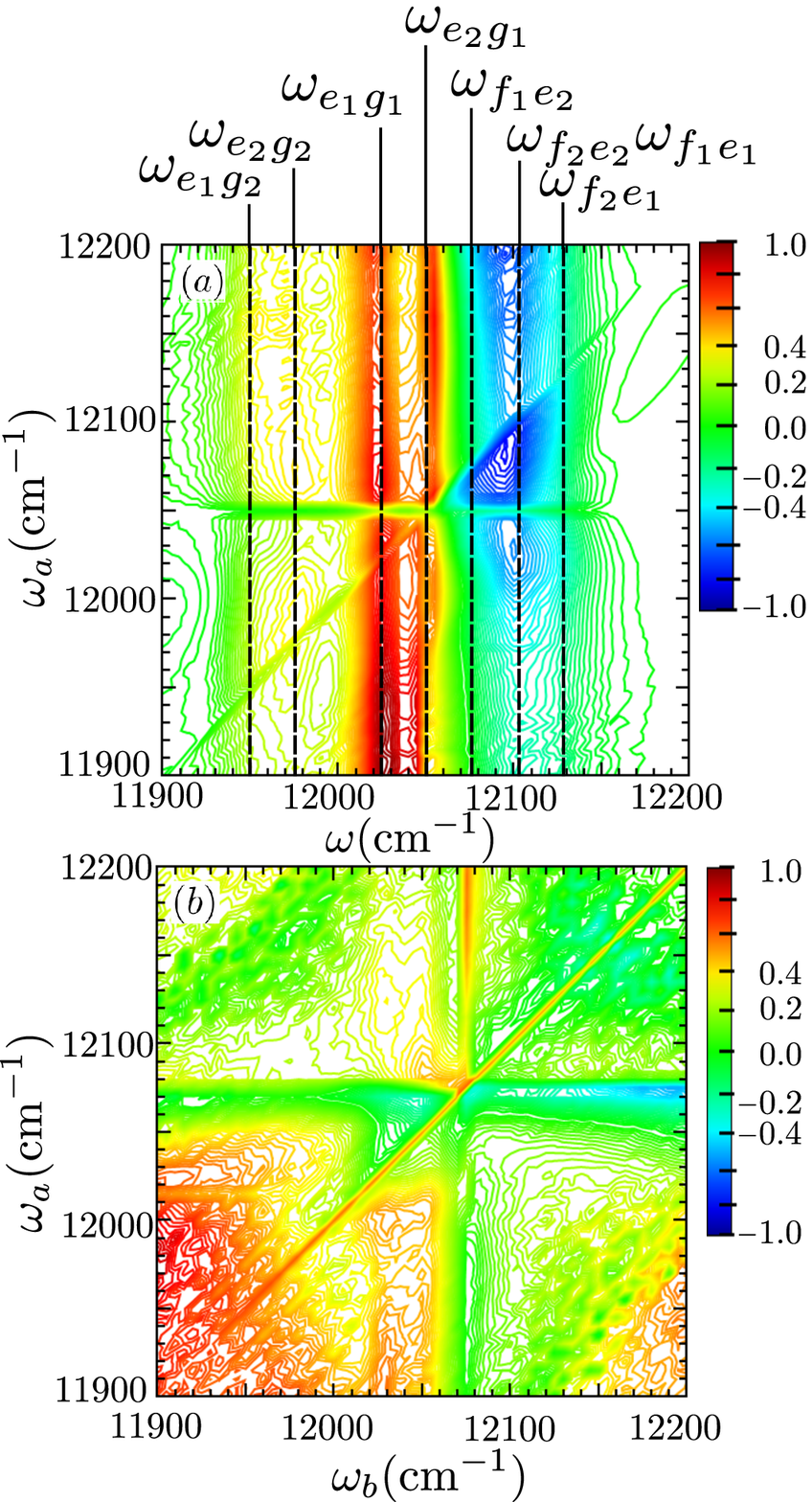}
\end{center}
\caption{(Color online) (a) The two-dimensional transmission signal, Eq. \eqref{wide}, is plotted for
  $\phi_2(\omega,\omega_a,\omega_{e_2g_1})$ with $\tau=-0.5\mathrm{cm}$.
(b) The 2D transmission spectrum with phase $\phi_2(\omega,\omega_a,\omega_b)$,
Eq. \eqref{Si}, vs phase-step positions
$\omega_a$ and $\omega_b$, is plotted for
$\tau=-0.5\mathrm{cm}$ and  $\omega=\omega_{f_2e_1}$
}\label{fig:2D_2_steps}
\end{figure}
The spectra contains diagonal peaks that extend above and below the diagonal line at $\omega=\omega_a$.
The main diagonal peaks are at $\omega=\omega_{e_1 g_1}$, $\omega_{e_2 g_1}$, $\omega_{f_1 e_1}$
and $\omega_{e_1 g_2}<\omega<\omega_{e_2 g_2}$. These peaks correspond to the peaks in Fig. \ref{fig:Linear}(b).
The phase-step changes the Rayleigh and Stokes peaks  from absorption to emission, and
the TPA peaks change from emission to absorption.
As seen in Fig. \ref{fig:Linear}(b), the Stokes peaks appear weaker than the other transition peaks in the spectra.
These peaks  appear above the diagonal line. They originate from the ladder diagrams (I)(a)(b) and (II)(a) in Fig. \ref{fig:overall}. Diagrams  (I)(b) and (II)(a) have the time sequence of arrows which alternate
in directions and diagram  (I)(a) has the time sequence of  two successive arrows with the same direction.
The $\omega=\omega_{e_1 g_1}$ peak extends above the diagonal line, while the $\omega=\omega_{e_2 g_1}$
extends below the diagonal line. The  main contribution for the  $\omega=\omega_{e_1 g_1}$  peak comes from ladder diagrams Fig. \ref{fig:overall}(I)(a) (b), (II)(a)(b). The dominant contribution is from diagrams (II)(a),(b) which
 have alternating time-ordering of the arrows direction. The  main contribution to $\omega=\omega_{e_2 g_1}$ peak is from the diagram (III)(b), which has two successive arrows in the same direction. The $\omega=\omega_{f_2 e_1}$, $\omega_{f_1 e_2}$,$\omega_{f_1 e_1}$ peaks appear below the diagonal line. These peaks can be traced back
to diagrams (I)(c), (II)(c), (III)(a) in Fig. \ref{fig:overall}, all of which have a time sequence of two successive arrows
in the same direction.
Overall, the peaks that extend below the diagonal line, have a dominate contribution from the diagrams which contain
time ordering of two successive arrows with the same direction and peaks that extend above the diagonal line have
a dominate contribution from the time sequence of arrows which alternate in direction.

In Fig. \ref{fig:appendix1} of Appendix B, we separate the transmission signal Fig. \ref{fig:2Dintense} into the  components $S_I(\omega,\omega_a)$, $S_{II}(\omega,\omega_a)$ and $S_{III}(\omega,\omega_a)$,
corresponding to  $\chi^{(3)}_I$, $\chi^{(3)}_{II}$, $\chi^{(3)}_{III}$ in Eq. \eqref{overallS}. The signal components,
show off-diagonal peaks, which appear when $\omega_a$ coincides with the transition frequencies. These peaks
are small compared to the diagonal peaks.
From Fig. \ref{fig:appendix1}, it can
be seen that the transmission signal with a phase-step is mostly composed of the
 $S_{II}(\omega,\omega_a)$ and $S_{III}(\omega,\omega_a)$ components.

The transmission signal with a negative phase-step  $\tau_a=-0.5 \mathrm{cm}$,  Eq. \eqref{diff}, is shown in Fig. \ref{fig:2Dintense}(b).
The diagonal  $\omega=\omega_{e_1 g_2}$, $\omega_{e_2 g_2}$ peaks are
much weaker, compared to the positive phase-step. However, the diagonal $\omega=\omega_{f_1 e_1}$, $\omega_{f_1 e_2}$, $\omega_{f_2 e_1}$  peaks
appear stronger, than with the positive phase-step.
These peaks originate from diagrams in  Fig. \ref{fig:overall}(I)(c), (II)(c), (III)(a). This shows that the negative going phase-step enhances
the TPA diagonal peaks stronger, whereas, a positive going phase-step enhances the $\omega=\omega_{e_1 g_2}$, $\omega_{e_2 g_2}$ peaks better.

 The difference of the transmission signal with a negative and positive phase-step
\begin{equation}
\Delta \tilde{S}_{\mathrm{step}}(\omega;\omega_a)=
S(\omega;  \phi_1(\omega,\omega_a,-\tau_a))-S(\omega; \phi_1(\omega,\omega_a,\tau_a))
\label{posneg}
\end{equation}
is plotted in Fig. \ref{fig:2Dintense}(c).
It
shows  peaks along the diagonal line. The spectral phase function
can be written as the sum of an even and odd function, $\exp(i\arctan(\pm x))=(1\pm x)/\sqrt{1+x^2}$.
The difference in the transmission signal with a positive and negative phase-step
gives the amplification by the odd part of the spectral phase function-$2x/\sqrt{1+x^2}$.
Figure \ref{fig:2Dintense}(c) confirms that the positive going phase-step enhances the
 $\omega=\omega_{e_1 g_2}$, $\omega_{e_2 g_2}$ peaks better than the negative
going, while the negative going phase-step enhances the $\omega=\omega_{f_1 e_1}$, $\omega_{f_1 e_2}$, $\omega_{f_2 e_1}$  peaks better than the positive. This effect is more clearly seen in the one-dimensional
transmission spectra $S(\omega; \phi_1(\omega,\omega_a))$,  Eq. \eqref{Si},   in  Fig. \ref{fig:Diff_steps}.
Positive step spectra are shown in the top row of Fig. \ref{fig:Diff_steps} for three phase-step positions.
The off-resonant phase-step position $\omega_a=11905 \mathrm{cm}^{-1}$ in (a)  shows that
the background is amplified. When the phase-step  coincides with a transition frequency
$\omega_a=\omega_{e_1g_2}$ in (b)  the peak becomes  amplified and the resolution is
increased.
This is  because the Greens functions,
$G_{\beta}(\omega)=(\omega-\omega_{\beta}+i\gamma)^{-1}$ in Eq. \eqref{S1}-\eqref{S3}
inverts the phase over a width
$\gamma$ about the resonance frequency $\omega=\omega_{\beta}$. The application
of the phase-step at $\omega_a=\omega_{\beta}$ inverts the phase again over a width $1/\tau_a$, enhancing the peak.
This was demonstrated by  Oron et al.\cite{oron_narrow-band_2002} in a CARS experiment.
Fig. \ref{fig:Diff_steps}(c) shows the enhancement of the TPA at $\omega_{f_1 e_2}$.
The negative phase-step is plotted in the bottom row of Fig.  \ref{fig:Diff_steps} for three values of the
phase-step position. For the phase-step off resonant, $\omega_a=11905 \mathrm{cm}^{-1}$, (d)
there is little amplification of the background. When $\omega_a=\omega_{e_1 g_2}$  in (e),
the enhancement becomes clear.  The negative step enhances the TPA $\omega_{f_1 e_2}$, in (f), better
than the Stokes peak. Fig. \ref{fig:Diff_steps}(b) shows that the positive step narrows and amplifies the Stokes peak
at $\omega_a=\omega_{e_1 g_2}$ better than the negative (e). However, the negative step in (f) enhances the TPA
at $\omega_a=\omega_{f_1 e_2}$ better than the positive and that this peak is enhanced down-ward
where as the $\omega_s=\omega_{e_1 g_2}$ peak is enhanced up-ward.

In this study, we only  considered homogeneous broadening.
Inhomogeneous broadening changes the line-shapes of the
peaks from diagrams (I) in Fig. \ref{fig:overall} from symmetric to asymmetric\cite{mukamelbook}.
The line-shapes from diagrams (II) and (III)
in Fig. \ref{fig:overall} remain symmetric with inhomogeneous broadening\cite{mukamelbook}. The transmission signal
with the phase-step is dominated by the transition peaks from diagrams (II) and (III) in Fig. \ref{fig:overall},
so we expect inhomogenous broadening to have little effect on the phase-step.

%
%
%
%
%
\section{ Two Dimensional Transmission signal with a phase-pulse}
%
%
%

We next present simulations which use a
$\pi$-phase-pulse spectral phase
\begin{equation}
\phi_2(\omega,\omega_a,\omega_b)= \arctan[\tau( \omega-\omega_a)]-\arctan[\tau( \omega-\omega_b)].
\label{arctan}
\end{equation}
The pulse lies between $\omega_a$ and $\omega_b$, with transition steepness $\tau$.
We first consider a narrow phase-pulse, with width close to the transition width of the step,
see Fig. \ref{fig:0}.
does not give a full $\pi$-inversion.  A full $\pi$-inversion is achieved with a wider phase-pulse,
$\omega_a-\omega_b=50 \mathrm{cm}^{-1}$ thin-red
line in Fig.  \ref{fig:0}.

The transmission spectrum for a narrow phase-pulse
\begin{equation}
\Delta S_{\mathrm{pulse}}(\omega;\tilde{\omega})=
S(\omega; \phi_2(\omega,
\tilde{\omega}+\frac{\Delta\omega}{2},
\tilde{\omega}-\frac{\Delta\omega}{2}, \tau))
-
S_0(\omega),
\label{diffpulse}
\end{equation}
 with width $\Delta\omega=10 \mathrm{cm}^{-1}$ and  position $\tilde{\omega}$
is displayed in Fig. \ref{fig:Pulse}(a) for a positive going pulse,  $\tau=0.5\mathrm{cm}$.
The vertical dashed-black lines mark the position of the transition peaks.
The positive going pulse in Fig. \ref{fig:Pulse}(a) has three vertically spread peaks at $\omega=\omega_{e_1 g_2}$,
$\omega_{e_1 g_1}$, $\omega_{f_1 e_2}$, corresponding to the three well pronounced peaks in Fig. \ref{fig:Linear}(b).
The effect of the phase-step changed the absorption
peaks in Fig. \ref{fig:Linear}(b) to emission peaks and vise versa. The effect of the phase-pulse gives the same
affect. The data in Fig. \ref{fig:Pulse}  is plotted on a scale, which gives the best visibility for peaks in the transmission spectrum.
 The main contribution to the transmission signal can be traced back to the diagrams (I) in Fig. \ref{fig:overall}.
See Fig. \ref{fig:appendix2} in the Appendix B for the individual $\Delta S^I_{\mathrm{pulse}}(\omega)$
 $\Delta  S^{II}_{\mathrm{pulse}}(\omega)$ and  $\Delta  S^{III}_{\mathrm{pulse}}(\omega)$
components. Signal component $\Delta S^{II}_{\mathrm{pulse}}(\omega,\omega_a)$  in Fig. \ref{fig:appendix2}  contributes two diagonal
peaks near $\omega=\omega_{e_2 g_1}$, $\omega_{f_1 e_1}$  and $\Delta S^{III}_{\mathrm{pulse}}(\omega,\omega_a)$  contribution is not noticeable.

The negative going phase-pulse  $\tau=-0.5\mathrm{cm}$ , Eq. \eqref{diffpulse}, is shown in Fig. \ref{fig:Pulse}(b).
The spectra are dominated by the contributions from diagrams (I) in Fig. \ref{fig:overall}.
The signal $\Delta S^{II}_{\mathrm{pulse}}(\omega,\omega_a)$ contributes two diagonal peaks at $\omega=\omega_{e_1g_1}$, $\omega_{f_1e_1}$
and $\Delta S^{III}_{\mathrm{pulse}}(\omega,\omega_a)$ contribution is not noticeable.
This shows that the positive or negative narrow phase-pulse can be used to suppress the contributions from
diagrams (II) and (III) in the transmission spectra. Note that with inhomogeneous broadening included
the signal with a phase-pulse should be significantly affected, since it is dominated by the diagrams (I).
  The negative pulse enhances the diagonal peak
near $\omega=\omega_{f_1 e_1}$ better than the positive pulse. It appears that the positive pulse
enhances the   $\omega=\omega_{e_1 g_2}$, $\omega_{e_2 g_2}$ better than the negative.

The difference in the transmission signal for a negative and positive pulse
\begin{equation}
\Delta {\tilde{S}}_{\mathrm{pulse}}(\omega;\tilde{\omega})=
S(\omega;\phi_2(\omega,
\tilde{\omega}+\frac{\Delta\omega}{2},
\tilde{\omega}-\frac{\Delta\omega}{2},- \tau))
-
S(\omega;\phi_2(\omega,
\tilde{\omega}+\frac{\Delta\omega}{2},
\tilde{\omega}-\frac{\Delta\omega}{2}, \tau))
\label{Diff2}
\end{equation}
is plotted in Fig. \ref{fig:Pulse}(c). The spectra is composed a diagonal peak near
$\omega=\omega_{f_1 e_1}$ and two small diagonal   peaks at
$\omega=\omega_{e_2 g_2}$, $\omega_{e_1 g_2}$.
The  two-dimensional spectra most resembles the Fig. \ref{fig:appendix2}(f)
from diagrams (II).

We have explored if a variable pulse width $\phi_2(\omega, \omega_a,\omega_b)$
with a negative-step located at $\omega_b=\omega_{e_2 g_1}$  could better enhance the
 peaks.
The $\omega=\omega_{e_2 g_1}$ peak appeared strong for a positive/negative
narrow-phase-pulse or step and always remained above the diagonal line.
It is interesting to see if there is any sensitivity to the phase-step position.
The transmission spectrum
\begin{equation}
\Delta \mathcal{S}_{\mathrm{pulse}}(\omega;\omega_a)=
S(\omega;\phi_2(\omega,\omega_a,\omega_b, \tau))
-
S_0(\omega).
\label{wide}
\end{equation}
with a variable pulse width is shown in Fig. \ref{fig:2D_2_steps}(a).
This creates a negative-going pulse for $\omega_a <\omega_{e_2 g_1}$
and positive-going pulse for $\omega_a  >\omega_{e_2 g_1}$.
The horizontal line at $\omega_a=\omega_b$ is where both phase-steps coincide and their effect vanishes.
The $\omega=\omega_{e_2 g_1}$ peak is well pronounced away from the diagonal line,
for any value of the pulse width.
Notice that the $\omega=\omega_{f_1 e_2}$, $\omega_{f_2 e_2}$ peaks
become strongly enhanced when the pulse changes from a negative to a positive going pulse.
Compare to the narrow-pulse Fig \ref{fig:Pulse}, the variable width enhances the
$\omega=\omega_{f_1 e_2}$, $\omega_{f_2 e_2}$ diagonal peaks stronger.
This is because the pulse is composed of a negative step located at $\omega_a$.

The phase-pulse has been used to enhance the TPA peaks $\omega_{f_i e_j}$ in the spectrum,
involving an intermediate resonant state\cite{dudovich_transform-limited_2001}.
 Two opposite phase-steps
located at the two transition frequencies, $\omega_{e_j g_1}$ and $\omega_{f_i e_j}$ involved in the TPA transition $\omega_{f_i g_1}$ where used.
If the intermediate state is located at $\omega_{f_i g_1}/2$, the phase-pulse would have zero width.
When the intermediate resonant state is de-tuned from $\omega_{f_i g_1}/2$ the phase-pulse has finite width
and can be employed to enhance the TPA peak.
We searched for a particular width that would
selectively enhance the TPA peak $\omega_{f_1 e_1}$ in the transmission spectrum.
The two-dimensional spectra, Eq. \eqref{Si},
with phase $\phi_2(\omega,\omega_a,\omega_b)$, are plotted in Fig. \ref{fig:2D_2_steps}(b) as a function of $\omega_a$ and $\omega_b$,
for $\omega=\omega_{f_1 e_2}$. The diagonal line at $\omega_a=\omega_b$ is where width of the phase-pulse is
zero. The phase-pulse is positive going for $\omega_b>\omega_a$ and negative going for $\omega_a>\omega_b$.
The diagonal peak  at $\omega_a=\omega_b=\omega_{f_1 e_2}$ shows that $\omega=\omega_{f_1 e_2}$ is optimized by employing a narrow negative pulse.
There is one horizontal line for $\omega_b>\omega_a$ and $\omega_a=\omega_{f_1 e_2}$. This is a wide pulse
with the negative going step at $\omega_a=\omega_{f_1 e_2}$. There is a vertical line for
$\omega_a>\omega_b$ and $\omega_b=\omega_{f_1 e_2}$, which is again a wide pulse with a negative step
located at $\omega_b=\omega_{f_1 e_2}$. Recall that in Fig. \ref{fig:2Dintense}(b) the transition peak $\omega_{f_1 e_2}$ was well enhanced using a negative step rather than a positive.
There are several regions where the background becomes pronounced: the bottom right and left corners.
For $\omega>\omega_{f_1e_2}$, the background is minimal. Overall, the spectra does not show any favorable
width for a wide pulse.

\begin{figure*}[h!t]
\begin{center}
\includegraphics[scale=0.3]{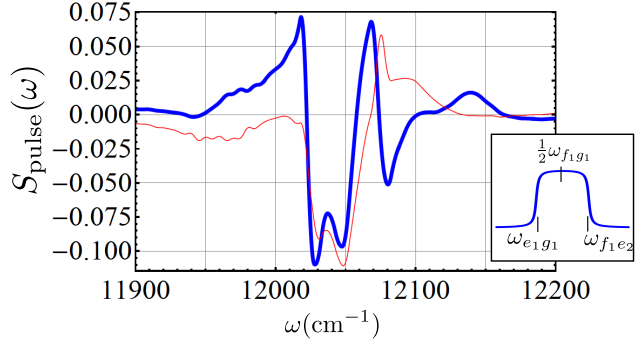}
\end{center}
\caption{(Color online) The  transmission signal Eq. \eqref{Si}
 $\phi_2(\omega,\omega_{e_1 g_1},\omega_{f_1 e_2})$
is plotted for a (thick-blue)  positive phase-pulse, $\tau=0.5\mathrm{cm}$
and a   (thin-red) negative phase-pulse $\tau=-0.5\mathrm{cm}$.
The width and position of the pulse used are shown in the inset.
}\label{fig:TPA}
\end{figure*}

We next chose, the width of the phase-pulse
to coincide with the transition frequencies  involved in the TPA transition $\omega_{f_1 g_1}$, with a phase-pulse centered at $\omega_{f_1 g_1}/2$. See the inset of Fig. \ref{fig:TPA}.
The transmission signal $S_{\mathrm{pulse}}(\omega)$, Eq. \eqref{Si},
is plotted  with a positive phase-pulse (blue-thick)  and a negative phase-pulse (thin-red).
Overall, the positive phase-pulse amplifies the TPA better than the negative phase-pulse.
This is because the positive phase-pulse is composed of a negative step at  $\omega_b=\omega_{f_1 e_2}$,
while the negative phase-pulse is composed of a positive step at $\omega_b=\omega_{f_1 e_2}$.

\begin{figure*}[h!t]
\begin{center}
\includegraphics[scale=0.3]{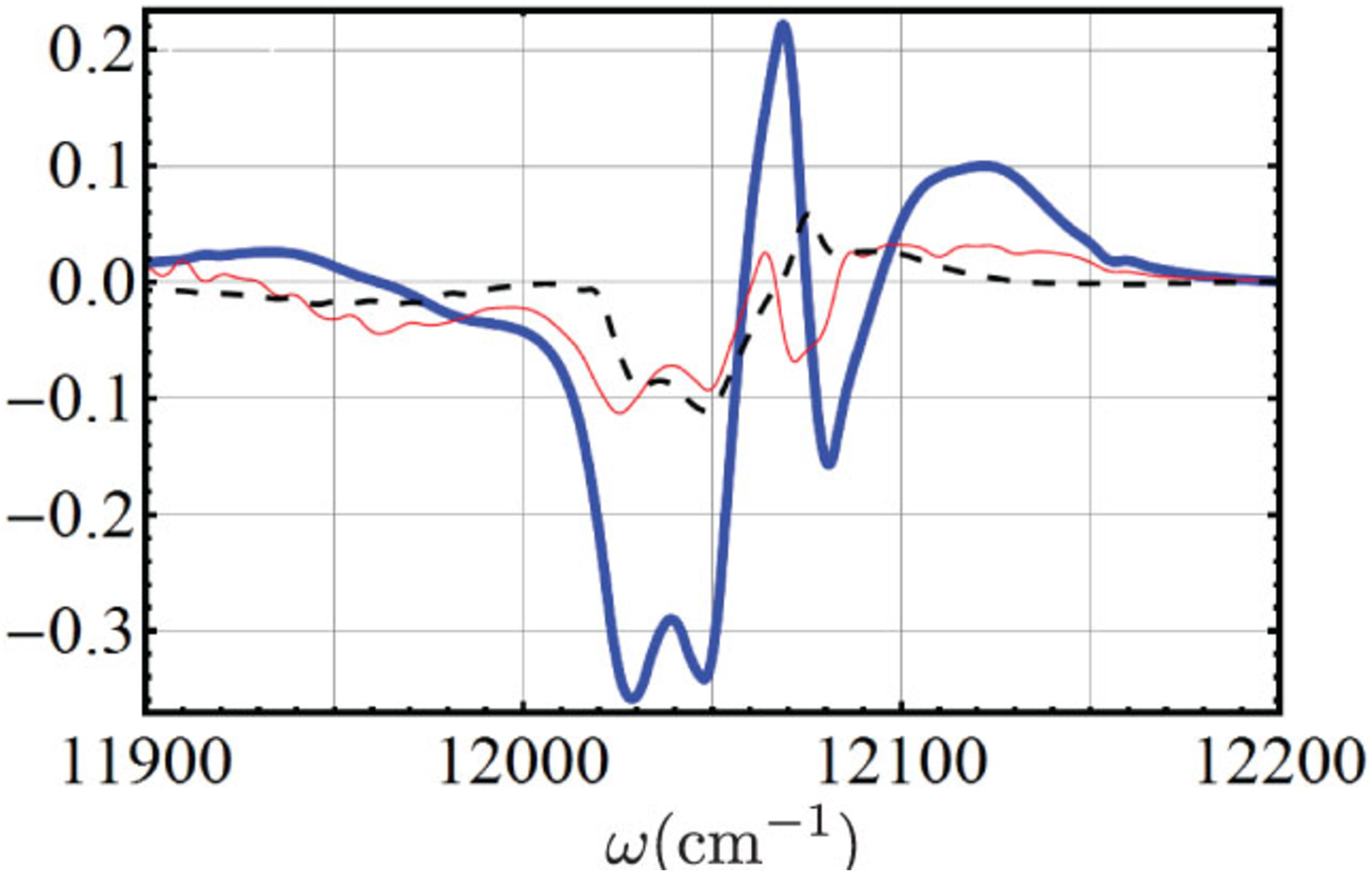}
\end{center}
\caption{(Color online) Three profiles for amplifying the peak
$\omega_{f_1e_2}$ in the transmission spectrum Eq. \eqref{Si} are compared; (thick-blue) phase-step,
$\phi_1(\omega,\omega_{f_1e_2})$; (thin-red) narrow-phase-step
$\phi_2(\omega,\omega_{f_1e_2}+\frac{1}{2}\Delta\omega,\omega_{f_1e_2}-\frac{1}{2}\Delta\omega)$, with
width $\Delta\omega=10\mathrm{cm}^{-1}$;
(dashed-black) wide-phase-pulse $\phi_2(\omega,\omega_{e_1 g_1},\omega_{f_1 e_2})$.
}\label{fig:compare}
\end{figure*}

In Fig. \ref{fig:compare}, we compare all three profiles for selectively enhancing the peak $\omega_{f_1 e_2}$.
The thick-blue line in  Fig. \ref{fig:compare} is the
transmission signal, Eq. \eqref{Si} with negative phase-step, $\phi_1(\omega,\omega_{f_1 e_2})$.
The thin-red line corresponds to the transmission signal,
Eq. \eqref{Si} with a negative narrow-phase-pulse
$\phi_2(\omega,\omega_{f_1 e_2}+\frac{1}{2}\Delta\omega,\omega_{f_1 e_2}-\frac{1}{2}\Delta\omega)$, with
width $\Delta\omega=10\mathrm{cm}^{-1}$.
The dashed-black line in
 the transmission signal is a positive wide-phase pulse $\phi_2(\omega,\omega_{e_1 g_1},\omega_{f_1e_2})$, shown in the inset of Fig. \ref{fig:TPA}.
Figure \ref{fig:compare} shows that the phase-step gives the best enhancement.
The phase-step centered at the resonant frequency inverts the phase to
compensate for the phase inversion of the Green's function and enhances the peak \cite{oron_narrow-band_2002}.
 The wide and narrow phase-pulses provide comparable enhancement.

\begin{figure*}[h!t]
\begin{center}
\includegraphics[scale=0.2]{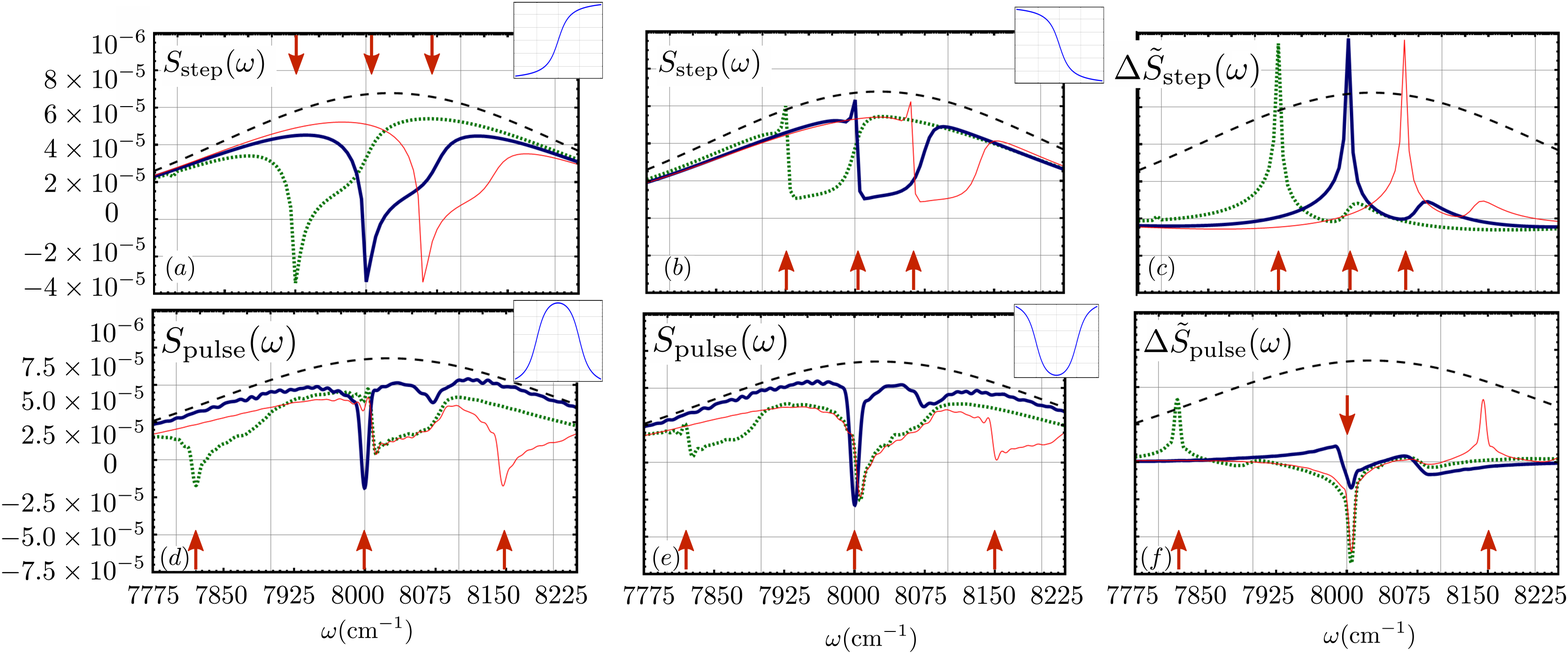}
\end{center}
\caption{(Color online)Top row: The  transmission signal Eq. \eqref{Si} with phase-step, $\phi_1(\omega,\omega_a)$, for three values of $\omega_a$:
 (dashed-green) $\omega_a=7925\mathrm{cm}^{-1}$,
(thick-blue) $\omega_a=\Omega_1$,
 (thin-red) $\omega_a=8060\mathrm{cm}^{-1}$.
(a) $\tau_{a}=0.5\mathrm{cm}$.
(b) $\tau_{a}=-0.5\mathrm{cm}$.
(c) Difference between (a) and (b), Eq. \eqref{posneg}.
Bottom row: The transmission signal with a phase-pulse
$ \phi_2(\omega, \tilde{\omega}+\frac{\Delta\omega}{2},
\tilde{\omega}-\frac{\Delta\omega}{2})$ with width  $\Delta\omega=10\mathrm{cm}^{-1}$
and three values of pulse position $\tilde{\omega}$:
(dashed-green) $\tilde{\omega}=7825\mathrm{cm}^{-1}$,
(thick-blue) $\tilde{\omega}=\Omega_1$,
(thin-red) $\tilde{\omega}=8150\mathrm{cm}^{-1}$,
(d)  $\tau=0.5\mathrm{cm}$.
(e)  $\tau=-0.5\mathrm{cm}$.
(f) Difference between (d) and (e), Eq. \eqref{Diff2}.
The dashed-black line is the transmission spectrum with no phase $\phi=0$.
 The red arrows marks the position of the phase-step.
The insets show the shape of the phase-step or pulse.
}\label{fig:raman}
\end{figure*}

%
%
%
%
%
\section{Off-resonant Stimulated Raman signals }
%
%
%

We now set the carrier frequency of the pulse to be off-resonant at $\Omega_1=8000\mathrm{cm}^{-1}$.
The transmission signal, Eq. \eqref{Si},
with a positive  phase-step is shown in Fig.
\ref{fig:raman}(a) for three values of $\omega_a$.
The black dashed line is the transmission signal, Eq. \eqref{S0} without pulse shaping, $\phi=0$.
With the phase-step, the transmission spectra shows a peak at $\omega=\omega_a$
with a width close to $\omega_{g_2 g_1}$.

The transmission signal with a negative phase-step,  Eq. \eqref{Si}, with $\tau_a=-0.5$, is shown in Fig. \ref{fig:raman}(b) for three values of $\omega_a$. The spectra show a peak at $\omega=\omega_a$
with width close $\omega_{g_2 g_1}$.
 The difference between the negative and positive-step, Eq. \eqref{posneg}, shows a peak at $\omega=\omega_a$ and a Stokes peak at $\omega_a+\omega_{g_2 g_1}$. We assume that
the molecule is initially in the $g_1$ state, meaning  that the anti-Stokes processes are not possible.

The transmission signal for a narrow positive phase-pulse Eq. \eqref{Si}
$S(\omega; \phi_2(\omega, \tilde{\omega}+\frac{\Delta\omega}{2},
\tilde{\omega}-\frac{\Delta\omega}{2}))$
with $\tau=0.5\mathrm{cm}$ is shown in Fig. \ref{fig:raman}(d) for three values of $\tilde{\omega}$.
The black-dashed line corresponds to the transmission spectrum Eq. \eqref{S0} with $\phi=0$.
The spectrum for $\tilde{\omega}=7825\mathrm{cm}^{-1}$, shows two peaks at $\Omega_1$ and $\tilde{\omega}$ with width close to $\omega_{g_2 g_1}$.
The same two peaks appear for $\tilde{\omega}=8150\mathrm{cm}^{-1}$.
When $\tilde{\omega}=\Omega_1$, there are two peaks at $\omega=\Omega_1$,
$\Omega_1+\omega_{g_2 g_1}$.
For a negative phase-step, $\tau=-0.5 \mathrm{cm}^{-1}$, we see the same peaks, with
slightly different line-shapes.

The
difference between the positive and negative phase-pulse, Eq. \eqref{Diff2} is shown in Fig. \ref{fig:raman}(f).
For $\tilde{\omega}=7825\mathrm{cm}^{-1}$, there are four peaks, $\omega=\tilde{\omega}$,
$\tilde{\omega}+\omega_{g_2 g_1}$, $\Omega_1$,  $\Omega_1+\omega_{g_2 g_1}$.
The same peaks occur for  $\tilde{\omega}=8150\mathrm{cm}^{-1}$. For
 $\tilde{\omega}=\Omega_1$ only two peaks occur  $\omega=\Omega_1$,  $\Omega_1+\omega_{g_2 g_1}$.
The Raman peaks at $\omega=\Omega_1 +\omega_{g_2 g_1}$, $\tilde{\omega}+\omega_{g_2 g_1}$
contain both absorption and emission features.
Note that there are some wiggles on the
 signal in Figs. \ref{fig:raman}(d), (e). The numerical value of the integrations calculated
 in Eq. \eqref{Si} is on the order $10^{-5}$, which is relatively small. The accuracy of the numerical integration
was set to $10^{-8}$.

%
%
%
%
%
\section{ Discussion}
%
%
%

We have simulated the nonlinear transmission signal of a broadband pulse with a phase-step and phase-pulse with finite-transition width.
Two dimensional plots of the transmission signal vs. the transmitted frequency and the position of a $\pi$-step
or pulse show diagonal peaks spread above and below the diagonal line.
The transmission spectra show that the phase-step or narrow-pulse can suppress
certain quantum pathways.
The diagonal peaks in the transmission spectra are sensitive to
the phase sign. The TPA
peaks are enhanced better for a negative step or narrow-pulse
rather than for a positive. The peaks from the Stokes process are
enhanced more for a positive step or narrow-pulse
compared to a negative. We found that
the sign of the
pulse or step
becomes relevant when the position of the step or pulse is close to the
transition peak frequencies.
A positive or negative phase-pulse selects particular peaks.
The narrow phase-pulse suppressed the contributions from the diagrams (II) and
(III) in Fig. \ref{fig:overall}, while the transmission spectra for the phase-step was dominated
by the diagrams (II) and (III) in Fig. \ref{fig:overall}.

The difference between a positive and negative phase-step or phase-pulse shows peaks along the diagonal line.
The $\arctan$ phase can expanded as the sum of an even and odd functions. Subtracting
the transmission signal with a positive and negative phase-step or pulses
gives the spectra which is enhanced by the odd-spectral phase.

We compared two protocols for enhancing the  TPA peaks in the transmission spectrum.
Refs. \cite{dudovich_transform-limited_2001} employed a wide pulse with the two phase-steps
 located at the two transition frequencies involved in the TPA. We plotted the two-dimensional transmission
spectra for a variable pulse width in Fig. \ref{fig:2D_2_steps}(b). However, the spectra showed that the TPA
is enhanced either for a narrow-phase-pulse or with a negative going phase-step.
The spectra did not show any peaks corresponding to a wide pulse with two phase-steps
 located at the two transition frequencies involved in the TPA.
 We compared this protocol to the phase-step and a narrow phase-pulse and showed that the step
gave the best amplification of the pulse.

\acknowledgments
We gratefully acknowledge the support of the Chemical Sciences,
Geosciences and Biosciences Division, Office of Basic Energy Sciences,
Office of Science, U.S. Department of Energy.  We also wish to
thank the National Science Foundation (Grant No. CHE-1058791).

\appendix
%
%
%
\section{Loop diagram expansion in $\chi^{(3)}$}
\label{sec:loop}
%
%
%
%

\begin{figure*}[t]
\begin{center}
\includegraphics[scale=0.6]{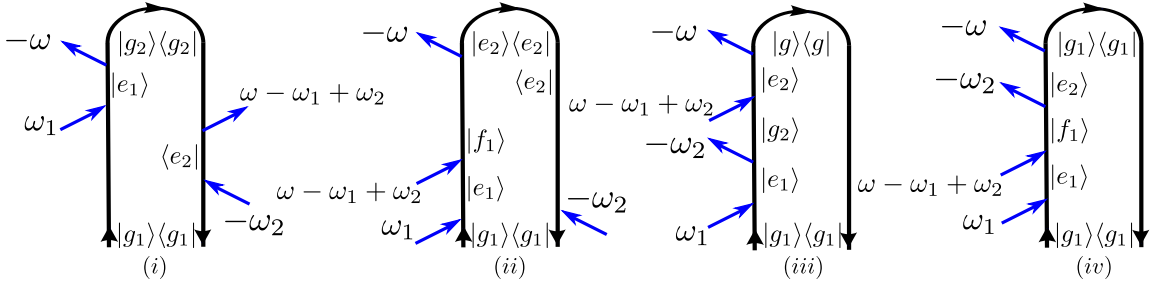}
\end{center}
\caption{(Color online)Loop diagrams for the frequency dispersed transmitted signal
signal  Eq. \eqref{tran} expanded to third-order in $H_{int}$.
The frequencies $\omega_1$ and $\omega_2$ correspond to the frequencies
$\omega_1$ and $\omega_2$ in Eq. \eqref{summation}.}
\label{fig:overall2}
\end{figure*}

The susceptibility Eq. \eqref{overallS} can be written in Hilbert space, corresponding the
loop diagrams in Fig. \ref{fig:overall2}, as
\begin{eqnarray}
&&
\chi^{(3)}(-\omega;\omega_1,-\omega_2,\omega-\omega_1+\omega_2)
=\chi^{(3)}_i(-\omega;\omega_1,-\omega_2,\omega-\omega_1+\omega_2)
+\chi^{(3)}_{ii}(-\omega;\omega_1,-\omega_2,\omega-\omega_1+\omega_2)
\nonumber\\&&
+\chi^{(3)}_{iii}(-\omega;\omega_1,-\omega_2,\omega-\omega_1+\omega_2)
+\chi^{(3)}_{iv}(-\omega;\omega_1,-\omega_2,\omega-\omega_1+\omega_2),
\label{summation}
\end{eqnarray}
where
\begin{widetext}
\begin{eqnarray}
\chi^{(3)}_{i}
(-\omega;\omega_1,-\omega_2,\omega-\omega_1+\omega_2)=&&
\left(\frac{-1}{2\pi \hbar}\right)^3
\sum_{g_i, e_i,f_i}
V_{g_1 e_2}
V_{ e_2 g_2}
V_{  g_2 e_1}
V_{ e_1 g_1}
G_{e_2}^\dagger(\omega_2)G_{g_2}^\dagger(-\omega+\omega_1)
G_{e_1}(\omega_1),
\nonumber\\&&
\label{App1}
\end{eqnarray}
\begin{eqnarray}
\chi^{(3)}_{ii}
(-\omega;\omega_1,-\omega_2,\omega-\omega_1+\omega_2)=&&
\left(\frac{-1}{2\pi \hbar}\right)^3
\sum_{g_i, e_i,f_i}
V_{g_1 e_2}
V_{ e_2 f_1}
V_{  f_1 e_1}
V_{ e_1 g_1}
G_{e_2}^\dagger(\omega_2)G_{f_1}(\omega+\omega_2)
G_{e_1}(\omega_1),
\nonumber\\&&
\label{App2}
\end{eqnarray}
\begin{eqnarray}
\chi^{(3)}_{iii}
(-\omega;\omega_1,-\omega_2,\omega-\omega_1+\omega_2)=&&
\left(\frac{-1}{2\pi \hbar}\right)^3
\sum_{g_i, e_i,f_i}
V_{g_1 e_2}
V_{ e_2 g_2}
V_{  g_2 e_1}
V_{ e_1 g_1}
G_{g_2}(\omega_1-\omega_2)G_{e_2}(\omega)
G_{e_1}(\omega_1),
\nonumber\\&&
\label{App3}
\end{eqnarray}
\begin{eqnarray}
\chi^{(3)}_{iv}
(-\omega;\omega_1,-\omega_2,\omega-\omega_1+\omega_2)=&&
\left(\frac{-1}{2\pi \hbar}\right)^3
\sum_{g_i, e_i,f_i}
V_{g_1 e_2}
V_{ e_2 f_1}
V_{  f_1 e_1}
V_{ e_1 g_1}
G_{e_2}(\omega)G_{f_1}(\omega+\omega_2)
G_{e_1}(\omega_1).
\nonumber\\&&
\label{App4}
\end{eqnarray}
\end{widetext}
Equation \eqref{overallS}  is similar to Eq. \eqref{summation}, with the dephasing rates $\gamma$ replaced
with the inverse lifetimes.

\section{Transmission spectra }

\begin{figure}[h!t]
\begin{center}
\includegraphics[scale=0.4]{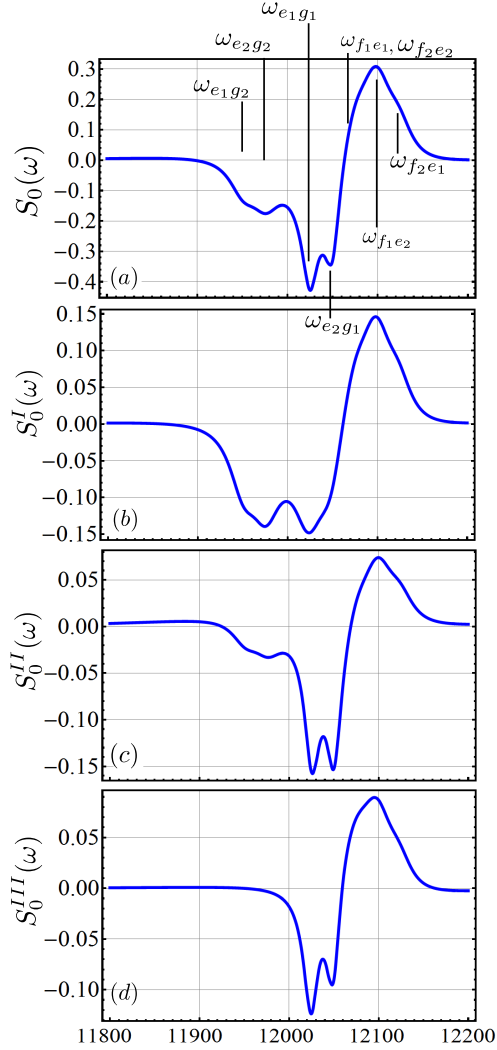}
\end{center}
\caption{(Color online)
(a) The frequency dispersed transmission signal
$S_0(\omega)$, Eq. \eqref{S0} using a Gaussian pulse Eq. \eqref{gaussian} for  $\phi=0$
$\sigma= 252 \mathrm{cm}^{-1}$ and $\Omega_1=12100 \mathrm{cm}^{-1}$. The components
 of $S_0(\omega)$
 (b) $S_0^I(\omega)$,
(c) $S_0^{II}(\omega)$,
(d) $S_0^{III}(\omega)$
corresponding to the susceptibility components,  Eq. \eqref{overallS}.
}
\label{fig:appendix0}
\end{figure}

The transmission spectrum $S_0(\omega)$ without pulse shaping is shown in Fig. \ref{fig:appendix0}(a).
The three  components $S^I_0(\omega)$, $S^{II}_0(\omega)$ and $S^{III}_0(\omega)$
of $S_0(\omega)$ are shown in Fig. \ref{fig:appendix0}(b)-(d), respectively. These components
correspond to the signal from diagrams (I), (II), (III) in Fig. \ref{fig:overall}.

\begin{figure}[h!]
\includegraphics[scale=0.50]{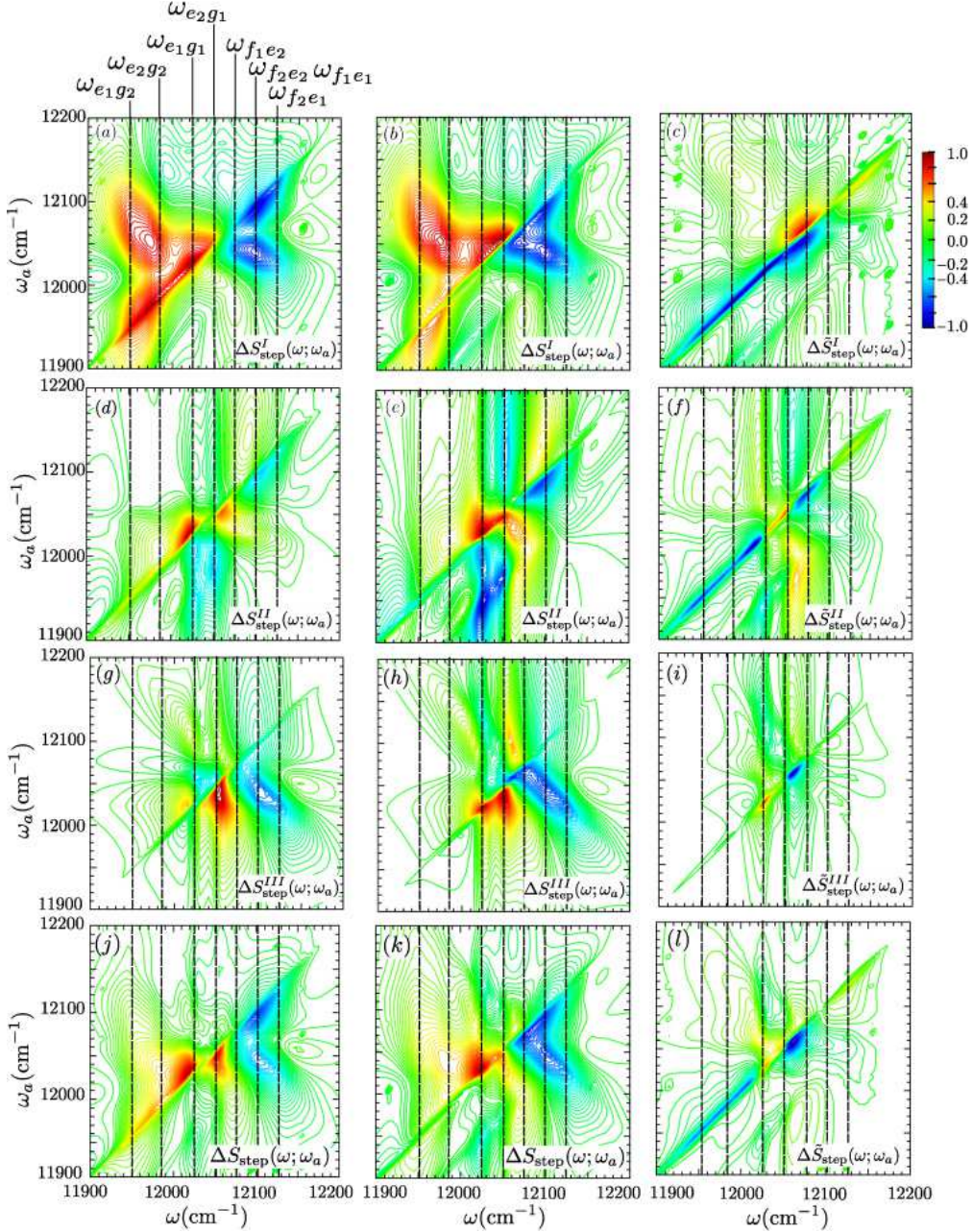}
\caption{(Color online) The two-dimensional  frequency dispersed difference-transmission signal
with phase $\phi_1(\omega,\omega_a)$, vs phase-step position, $\omega_a$.
(Left column) Positive step, $\Delta S^i_{\mathrm{step}}(\omega,\omega_a)$,
  Eq. \eqref{adiff} with $\tau_a=0.5 \mathrm{cm}$;
(Middle column) negative step, $\Delta S^i_{\mathrm{step}}(\omega,\omega_a)$,
  Eq. \eqref{adiff} with $\tau_a=-0.5 \mathrm{cm}$;
(Right column) difference between a positive and negative step,
$\Delta \tilde{S}^i_{\mathrm{step}}(\omega,\omega_a)$
 Eq. \eqref{aposneg}.
 (First row) $ S^I_{\mathrm{step}}$;
(Second row) $ S^{II}_{\mathrm{step}}$;
(Third row) $S^{III}_{\mathrm{step}}$;
(Fourth row)  the total transmission signal, (j), (k) Eq. \eqref{total1}; (l)
Eq. \eqref{total2}
The vertical dashed-black lines mark the transition frequencies.
}
\label{fig:appendix1}
\end{figure}

The  two-dimensional  frequency dispersed transmission signal
\begin{equation}
\Delta S^i_{\mathrm{step}}(\omega;\omega_a)=
S^i(\omega; \phi_1(\omega,\omega_a,\tau_a))
-
S^i_0(\omega),
\label{adiff}
\end{equation}
 is plotted in the first two columns of Fig. \ref{fig:appendix1}. The index $i=I,II,III$ represents
the signal from diagrams (I), (II), (III) in Fig. \ref{fig:overall},  respectively. The first column corresponds to a positive phase-step  $\tau_a=0.5 \mathrm{cm}$ and
the second column to a negative phase-step  $\tau_a=-0.5 \mathrm{cm}$. The difference  transmission signal with a negative and positive phase-step
\begin{equation}
\Delta \tilde{S}^i_{\mathrm{step}}(\omega;\omega_a)=
S(\omega;  \phi_1(\omega,\omega_a,-\tau_a))-S(\omega; \phi_1(\omega,\omega_a,\tau_a))
\label{aposneg}
\end{equation}
is plotted in the third column of Fig. \ref{fig:appendix1}.
The vertical black-dashed  lines mark the positions of the
transition peaks.
The first three rows corresponds the transmission signal components from the ladder diagrams in Fig. \ref{fig:overall}(I), (II), (III), respectively. The fourth row corresponds to the total transmission signal.
The sum
\begin{equation}
\Delta S_{\mathrm{step}}(\omega;\omega_a)
=
\Delta S^I_{\mathrm{step}}(\omega;\omega_a)
+\Delta S^{II}_{\mathrm{step}}(\omega;\omega_a)
+\Delta S^{III}_{\mathrm{step}}(\omega;\omega_a)
\label{total1}
\end{equation}
is plotted Figs. \ref{fig:appendix1}(j) (k), for a positive and negative step, respectively.
The difference between (j) and (k),
or the sum
\begin{equation}
\Delta  \tilde{S}_{\mathrm{step}}(\omega;\omega_a)
=
\Delta  \tilde{S}^I_{\mathrm{step}}(\omega;\omega_a)
+\Delta  \tilde{S}^{II}_{\mathrm{step}}(\omega;\omega_a)
+\Delta  \tilde{S}^{III}_{\mathrm{step}}(\omega;\omega_a)
\label{total2}
\end{equation}
 is plotted in Fig. \ref{fig:appendix1}(l).

\begin{figure}[h!]
\includegraphics[scale=0.50]{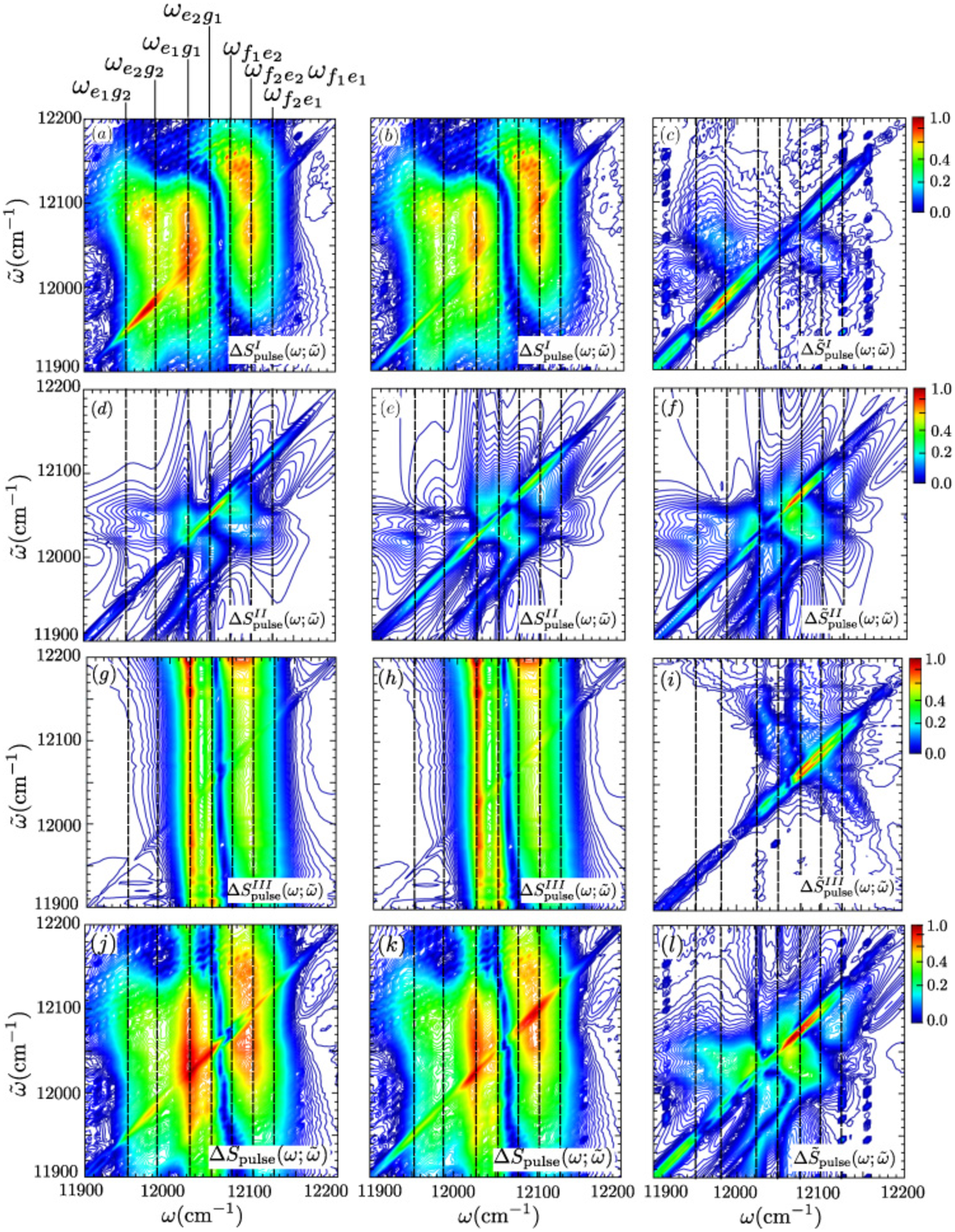}
\caption{(Color online)  The two-dimensional  frequency dispersed difference-transmission signal
vs the phase-pulse, $\phi_2(\omega,\tilde{\omega}+\frac{\Delta \omega}{2},
\tilde{\omega}-\frac{\Delta \omega}{2})$, position $\tilde{\omega}$, with width $\Delta\omega=10 \mathrm{cm}^{-1}$.
(Left column) Positive pulse, $\Delta S^i_{\mathrm{pulse}}(\omega,\tilde{\omega})$;
  Eq. \eqref{adiffpulse} with $\tau=0.5 \mathrm{cm}$,
(Middle column) negative pulse, $\Delta S^i_{\mathrm{pulse}}(\omega,\tilde{\omega})$;
  Eq. \eqref{adiffpulse} with $\tau=-0.5 \mathrm{cm}$;
(Right column) difference between a positive and negative pulse,
$\Delta \tilde{S}^i_{\mathrm{pulse}}(\omega,\tilde{\omega})$
 Eq. \eqref{aDiff2}.
 (First row) $ S^I_{\mathrm{pulse}}$;
(Second row) $ S^{II}_{\mathrm{pulse}}$;
(Third row) $S^{III}_{\mathrm{pulse}}$;
(Fourth row)  the total transmission signal, (j), (k) Eq. \eqref{total3}; (l)
Eq. \eqref{total4}.
The vertical dashed-black lines mark the transition frequencies.
}
\label{fig:appendix2}
\end{figure}

The difference transmission spectrum for a narrow phase-pulse  with width $\Delta\omega=10 \mathrm{cm}^{-1}$ is plotted  in Fig. \ref{fig:appendix2}.  The difference
\begin{equation}
\Delta S^{i}_{\mathrm{pulse}}(\omega;\tilde{\omega})=
S^{i}(\omega; \phi_2(\omega,\tilde{\omega}+\frac{\Delta \omega}{2},
\tilde{\omega}-\frac{\Delta \omega}{2}, \tau))
-
S^{i}_0(\omega),
\label{adiffpulse}
\end{equation}
is displayed in the first two columns of Fig. \ref{fig:appendix2},
where the index $i$ can be $I$, $II$, or $III$.
The first and second column corresponds to a positive and negative pulse, respectively.
The difference in the transmission signal with a negative and positive pulse
\begin{equation}
\Delta {\tilde{S}}^i_{\mathrm{pulse}}(\omega;\tilde{\omega})=
S^i(\omega;\phi_2(\omega,\tilde{\omega}+\frac{\Delta \omega}{2},
\tilde{\omega}-\frac{\Delta \omega}{2},- \tau))
-
S^i(\omega;\phi_2(\omega,\tilde{\omega}+\frac{\Delta \omega}{2},
\tilde{\omega}-\frac{\Delta \omega}{2}, \tau))
\label{aDiff2}
\end{equation}
is plotted in the third column of Fig. \ref{fig:appendix2}.
The sum
\begin{equation}
\Delta S_{\mathrm{pulse}}(\omega;\omega_a)
=
\Delta S^I_{\mathrm{pulse}}(\omega;\omega_a)
+\Delta S^{II}_{\mathrm{pulse}}(\omega;\omega_a)
+\Delta S^{III}_{\mathrm{pulse}}(\omega;\omega_a)
\label{total3}
\end{equation}
is plotted Figs. \ref{fig:appendix2}(j) (k), for a positive and negative step, respectively.
The difference between (j) and (k),
or the sum
\begin{equation}
\Delta  \tilde{S}_{\mathrm{pulse}}(\omega;\omega_a)
=
\Delta  \tilde{S}^I_{\mathrm{pulse}}(\omega;\omega_a)
+\Delta  \tilde{S}^{II}_{\mathrm{pulse}}(\omega;\omega_a)
+\Delta  \tilde{S}^{III}_{\mathrm{pulse}}(\omega;\omega_a)
\label{total4}
\end{equation}
 is plotted in Fig. \ref{fig:appendix2}(l).

\end{document}